\documentstyle[12pt,aasms4]{article}
\def\be{\begin{equation}}
\def\ee{\end{equation}}
\def\ba{\begin{eqnarray}}
\def\ea{\end{eqnarray}}
\def\etal{et al.}
\def\bfr{{\bf r}}

\def\ga{\mathrel{\mathpalette\fun >}}
\def\fun#1#2{\lower3.6pt\vbox{\baselineskip0pt\lineskip.9pt
        \ialign{$\mathsurround=0pt#1\hfill##\hfil$\crcr#2\crcr\sim\crcr}}}

\slugcomment{To be submitted to ApJ}

\begin{document}
\null\vspace{-62pt}
\begin{flushright}
revised version (July 27, 1998)
\end{flushright}

\title{ 
Cosmology in the Next Millennium: 
Combining \\MAP and SDSS Data
to Constrain Inflationary Models}
\author{Yun Wang, David N. Spergel, \& Michael A. Strauss\footnote{Alfred P. Sloan Foundation Fellow}$^,$\footnote{Cottrell Scholar of Research Corporation}}
\affil{{\it Princeton University Observatory} \\
{\it Peyton Hall, Princeton, NJ 08544\\}
{\it email: ywang,dns,strauss@astro.princeton.edu}}

\vspace{.4in}
\centerline{\bf Abstract}
\begin{quotation}

The existence of primordial adiabatic Gaussian random-phase density 
fluctuations is a generic prediction of inflation. The properties of 
these fluctuations are completely specified by their power spectrum 
$A_S^2(k)$. The basic cosmological parameters and the primordial power 
spectrum together completely specify predictions for the cosmic microwave 
background radiation anisotropy and large scale structure.
Here we show how we can strongly constrain both $A_S^2(k)$ and the 
cosmological parameters by combining the data from the Microwave 
Anisotropy Probe (MAP) and the galaxy redshift survey from the Sloan 
Digital Sky Survey (SDSS). We allow $A_S^2(k)$ to be a free function, 
and thus probe features in the primordial power spectrum on all scales.
If we assume that the cosmological parameters are known {\it a priori}
and that galaxy bias is linear and scale-independent, and 
neglect non-linear redshift distortions, 
the primordial power spectrum in 20 steps in $\log k$ to 
$k\leq 0.5h$Mpc$^{-1}$ can be determined to $\sim 16\%$  accuracy 
for $k\sim 0.01h$Mpc$^{-1}$, and to $\sim  1\%$ accuracy for $k\sim 
0.1h$Mpc$^{-1}$. The uncertainty in the primordial power spectrum 
increases by a factor up to 3 on small scales if we solve simultaneously 
for the dimensionless Hubble constant $h$, the cosmological constant 
${\Lambda}$, the baryon fraction $\Omega_b$, the reionization optical 
depth $\tau_{ri}$, and the effective bias between the matter density field 
and the redshift space galaxy density field $b_{\it eff}$.
Alternately, if we restrict $A_S^2(k)$ to be a power law, we find that
inclusion of the SDSS data breaks the degeneracy between the amplitude
of the power spectrum and the optical depth inherent in the MAP data,
significantly reduces the uncertainties in the determination of the 
matter density and the cosmological constant, and allows a determination 
of the galaxy bias parameter. 
Thus, combining the MAP and SDSS data allows the independent measurement
of important cosmological parameters, and a 
measurement of the primordial power spectrum independent of
inflationary models, giving us valuable
information on physics in the early Universe, and providing clues to
the correct inflationary model.

\end{quotation}


\section{Introduction}

Standard cosmology poses three basic conundrums: 
why the universe is homogeneous on large scales, 
why the cosmological density parameter $\Omega_m$ is close to unity, 
and what seeds the large-scale structure in the universe.
Inflation provides compelling solutions to all these 
problems (\cite{Kolb&Turner}).

The cosmic microwave background radiation (CMB) is observed to be isotropic 
to an accuracy of better than $10^{-4}$. 
This implies that the Universe
was homogeneous to this accuracy over many 
horizon scales at last scattering. 
Within the standard cosmology there is no causal process
which could have created the observed homogeneity. The basic idea of
inflation is that the observable Universe grew from an initial patch 
small enough to fit inside the horizon. 
It is hypothesized that there were particle physics mechanisms 
which led to
the rapid expansion of the early Universe required to produce 
the huge growth factor necessary to solve the cosmological problems
above. 
In this paper, we assume that inflation produces
primordial adiabatic Gaussian random-phase density 
fluctuations due to quantum fluctuations in the inflaton field.
The properties of these fluctuations are completely specified
by their power spectrum $A_S^2(k)$\footnote{Certain two-field inflationary models predict isocurvature
as well as adiabatic fluctuations (\cite{Kof87}).  Since
the CMB and LSS predictions of isocurvature models differ
significantly from adiabatic models and since these differences
are not degenerate with parameter variation, we could, in principle,
also fit for an isocurvature power spectrum.  However, in this
paper, 
we limit ourselves
to considering adiabatic fluctuations.}.  This paper will emphasize
how well the shape of the primordial power spectrum can be measured
from CMB anisotropy and redshift survey data. 

There currently exists a broad range of inflationary models
(\cite{Kolb96,Turner97}).
Some of these models predict power spectra that are
almost exactly scale-invariant (\cite{Linde83}),
or are described by a power law with spectral index less than one
(\cite{naturalinf,extendedinf}), 
while others predict power spectra
with slowly varying spectral indices (\cite{Wang94}), or
with broken scale invariance (\cite{Holman91ab,Adams97,Les97}).

Indeed, there are physical reasons to believe that the primordial power
spectrum has breaks in its power-law form. 
In effective (i.e., not complete)
theories with two scalar fields, 
inflation may occur in two stages.
The two stages of inflation can generate
density perturbations with different power-law indices, with
a step in amplitude of the primordial power
spectrum on the scale of the transition
between the two phases (\cite{Holman91ab}).
Fry \& Wang (1992) found that such models can have a significant 
signature on small-scale CMB anisotropies.

Recently, Adams, Ross, \& Sarkar (1997) proposed a new
multiple inflation model. Since attempts at an unified description
of the strong, weak and electromagnetic interactions usually involve
several stages of spontaneous symmetry breaking, they considered the
effects of such symmetry breaking during an era of inflation in 
supergravity models. They found that there can be a succession
of short bursts of inflation; the density perturbations produced during
each burst is nearly scale-invariant but with differing amplitudes,
and between each burst there is a brief period during which scale-invariance
is badly broken.

Given the range of possibilities for the primordial power spectrum,
we would like to quantify how well it can be measured generically.
Thus, we take the primordial power spectrum to be a free function
in this paper.

The upcoming data from the Microwave Anisotropy Probe (MAP; cf., Bennett
\etal\ 1997; {\tt http://map.gsfc.nasa.gov}) and
the Sloan Digital Sky Survey (SDSS; cf., Gunn \& Weinberg 1996)
provide a unique opportunity for constraining inflationary models.
It is useful to expand the temperature fluctuations in the CMB
into spherical harmonics: 
$\delta T/T (\hat{\bf r})= \sum_{l,m} a_{T,lm} Y_{lm}(\hat{\bf r})$,
where $\hat{\bf r}$ is the unit direction vector in the sky.
MAP measures the angular power spectrum (\cite{SeljakZ96})
\be
\label{C_l def}
C_{Tl} \equiv \langle |a_{T,lm}|^2 \rangle= (4\pi)^2 \int \frac{dk}{k}\,
A_S^2(k)\, \left| \Delta_{Tl} (k, \tau=\tau_0) \right|^2,
\ee
where $A_S^2(k)$ is the power spectrum of the primordial density 
fluctuations (defined such that $A_S^2(k)=1$ for the scale-invariant
Harrison-Zel'dovich  
spectrum), $\Delta_{Tl} (k, \tau=\tau_0)$ is an integral over 
conformal time $\tau$ of the sources which generate the CMB fluctuations,
and $\tau_0$ is the conformal time today.

The power spectrum of mass fluctuations in the linear regime today is
\be
P(k)=P_0 \,k A_S^2(k)\, T^2(k),
\ee
where $P_0$ is a normalization constant, and $T(k)$ is the transfer 
function, which depends on physics at matter-radiation equality
and decoupling. Note that $P_0$ has been absorbed into the
definition of $\Delta_{Tl} (k, \tau=\tau_0)$ in Eq.~(\ref{C_l def}).

The galaxy redshift survey from the SDSS (cf., Strauss 1997 for a
description) will allow a determination of $P_G(k)$,
the galaxy power spectrum in redshift space.  
The quantity $P_G(k)$ differs from the
mass power spectrum $P(k)$ due to two effects.  First, the galaxy
distribution may be biased with respect to the mass distribution. On
large scales, models (Weinberg 1995; Kauffmann, Nusser, \& Steinmetz
1997; Scherrer \& Weinberg 1997) indicate that the mass and galaxy
density fields are directly proportional to one another.   
The proportionality constant is referred
to as the galaxy bias parameter, $b$.  Second, peculiar
velocities cause the density contrast of galaxies in redshift space to
appear appreciably stronger than in real space.  In linear theory, the
power spectrum is boosted by a factor $1 + {2 \over 3} \beta + {1
\over 5} \beta^2$, where $\beta \equiv \Omega_m^{0.6}/b$
($\Omega_m$ is the matter density in units of the 
critical density of the Universe $\rho_c \equiv 3H_0^2/(8\pi G)$,
with $H_0$ denoting the Hubble constant).
The net result (Kaiser 1987; cf., Hamilton 1998 for a review)
is that on linear scales, the galaxy power spectrum is given by:
\begin{equation} 
\label{eq:PG}
P_G(k) = b^2 \left(1 +  {2 \over 3} \beta + {1 \over 5} \beta^2\right)
P(k) \equiv b_{\it eff}^2 P(k). 
\end{equation}

The observables $C_{Tl}$ and $P_G(k)$ depend on the cosmological
parameters $H_0$, $\Omega_b$ (baryon density in units of $\rho_c$),
$\Omega_m$, $\Omega_{\Lambda}$ (density contribution from the
Cosmological Constant in units of $\rho_c$), and $\tau_{ri}$
(reionization optical depth) only through $\Delta_{Tl} (k,
\tau=\tau_0)$, $b_{\it eff}$, and $T(k)$. The MAP data alone will
allow rather tight constraints on the baryon/photon ratio (which
determines $\Omega_b h^2$), the matter/photon ratio (which determines
$\Omega_m h^2$), and the geometry of the universe, and test the basic
inflationary scenario (Spergel 1994; Knox 1995; Jungman \etal\ 1996; Bond,
Efstathiou \& Tegmark 1997; Zaldarriaga, Spergel \& Seljak 1997;
Dodelson, Kinney \& Kolb 1997; Lidsey \etal\ 1997; Copeland, Grivell
\& Liddle 1997; Efstathiou \& Bond 1998; Eisenstein, Hu \& Tegmark 1998). 
However, there is a near-degeneracy in several sets
of parameters (Bond \etal\ 1994; Bond, Efstathiou \& Tegmark 1997;
Zaldarriaga, Spergel \& Seljak 1997; \cite{HD97}) 
including the overall amplitude of the power spectrum at $k \sim 0.1\,h\rm Mpc^{-1}$ and the optical depth.  Since
$\Delta_{Tl} (k, \tau=\tau_0)$ and $T(k)$ depend on the cosmological
parameters differently, by combining the MAP and SDSS data, we can
break these degeneracies and determine cosmological parameters to
impressive accuracies.  Several other recent papers have also considered
how well cosmological parameters can be constrained with the
combination of future CMB and galaxy survey data: Tegmark
(1997) has studied a somewhat different and smaller set of the cosmological
parameters than we consider here; Hu, Eisenstein, \& Tegmark (1997) 
and Eisenstein, Hu, \& Tegmark (1998)
have tested the sensitivity of the data to the neutrino mass, 
which we do not consider in this paper.  See also Lineweaver \& Barbosa (1998)
and Webster \etal\ (1998), who fit for cosmological
parameters using the best available CMB and redshift data available.

Both $C_{Tl}$ and $P_G(k)$ are proportional to the amplitude of
the power spectrum of the primordial density fluctuations $A_S^2(k)$.
Because of the finite resolution of the MAP satellite, the errors on
the $C_{Tl}$ increase rapidly with wavenumber for $l \ga 800$,
while the errors on $P_G(k)$ from the SDSS decrease with wavenumber; 
MAP and SDSS thus quite naturally complement each other in
the determination of $A_S^2(k)$.
Assuming that inflation occurred, all the obtainable information
about the inflationary model is contained in $A_S^2(k)$. 

Some inflationary models predict a significant gravity wave
contribution to the microwave background fluctuation spectrum
at small $l$.  However, even  in these models, gravity
waves do not make a significant contribution to the microwave
background fluctuation spectrum at $l \ga 100$ (\cite{Crit93,Dode94,Allen95,Wang96,Turner96}).
Since most of the information on the cosmological parameters 
comes from relatively small angular scales ($l \ga 100$),
the inclusion of tensor perturbations would not change our results 
qualitatively. 

We parameterize $A_S^2(k)$ by a series of steps equally spaced in
$\log k$ bins, 
each step with a constant amplitude in its bin. By taking these
amplitudes to be independent parameters, we probe features in the  
primordial power spectrum on all scales.
The measurement of the primordial power spectrum independent of
specific inflationary models 
from the combined MAP and SDSS data should shed light on our
understanding of the early Universe.

In Section 2, we present the statistical technique we use in
estimating errors of parameters.  Before considering a general
primordial power spectrum, we determine the accuracy with which
cosmological parameters will be measured with a pure power law
$A_S^2(k)$ in Section 3. 
In Section 4, we relax this constraint, and discuss the determination
of the primordial power spectrum, using the stepwise parameterization
described above. Section 5 contains a summary of the paper.

\section{The Fisher Information Matrix}

The Fisher information matrix of a given set of parameters, ${\bf s}$,
approximately quantifies the amount of
information on ${\bf s}$ that we ``expect'' to get from our future data.
The Fisher matrix can be written as
\be
\label{eq:fisherdef}
F_{ij}=- \frac{\partial ^2 \ln\,L}{\partial s_i \partial s_j},
\ee
where $L$ is the likelihood function, 
the expected probability distribution of the observables given
parameters ${\bf s}$.  
The Cram{\'e}r-Rao inequality (\cite{Cramer-Rao}) states that
no unbiased method can measure the $i$-th parameter with standard
deviation less than $1/\sqrt{F_{ii}}$ if other parameters are known,
and less than $\sqrt{ ({\bf F}^{-1})_{ii} }\,$ if other parameters
are estimated from the data as well.
Note that the derivatives in Equation (\ref{eq:fisherdef})
are calculated assuming that the cosmological parameters are given by
an a priori model, and thus the errors on the parameters are somewhat 
dependent on the assumed model.

The likelihood function $L$ in principle can be that of the raw data;
see, for example, the discussion in Tegmark \etal~(1998) and
Dodelson, Hui, \& Jaffe (1997).  But as Tegmark, Taylor, \& Heavens
(1997) describe, it is far more convenient to work with the data in a
more compressed form.  For the galaxy data, we work with the observed
galaxy power spectrum, which can be determined using the techniques
reviewed, e.g., in Strauss (1997) or Tegmark \etal~(1998).  The
values of the measured power spectrum are approximately
statistically independent if measured at values of
$k$ spaced at intervals of $2\,\pi/R$, where
$R$ is the characteristic length of the survey volume.   Similarly,
for a full-sky CMB mapping experiment like MAP, the individual $C_l$'s are
very nearly statistically independent (Oh, Spergel \& Hinshaw 1998). 

\subsection{CMB Data}

MAP will measure the CMB polarization as well as temperature anisotropy.
The CMB radiation field is described by a 2$\times$2 intensity tensor 
$I_{ij}$. The Stokes parameters are
$Q\equiv (I_{11}-I_{22})/4$ and $U\equiv I_{12}/2$,
while the temperature anisotropy is $\delta T/T \equiv (I_{11}+I_{22})/4$.
The temperature anisotropy can be expanded in standard spherical harmonics,
while $(Q\pm iU)$ have to be expanded using spin weight $\pm$2 
spherical harmonics $_{\pm 2} Y_{lm}$ 
(\cite{Gelfand63,Goldberg67,ZS97}; see also \cite{Kamion97}
for an alternative expansion):
\ba
\delta T/T (\hat{\bf n}) &=& \sum_{lm} a_{T,lm} Y_{lm} (\hat{\bf n}), 
\nonumber\\
(Q+iU) (\hat{\bf n}) &= &\sum_{lm} a_{2,lm} \, _2 Y_{lm} (\hat{\bf n}), \nonumber\\
(Q-iU) (\hat{\bf n}) &= &\sum_{lm} a_{-2,lm} \, _{-2} Y_{lm} (\hat{\bf n}),
\ea
where $\hat{\bf n}$ is the unit direction vector in the sky.
The even and odd parity linear combinations of $a_{2,lm}$ and $a_{-2,lm}$ are
\ba
a_{E,lm}&=& -(a_{2,lm}+a_{-2,lm})/2, \nonumber\\
a_{B,lm}&= &i(a_{2,lm}-a_{-2,lm})/2.
\ea
Note that $a_{E,lm}$ and $a_{B,lm}$ behave analogously to
the electric and magnetic fields respectively under parity transformation.
The statistics of the CMB are specified by the power spectra
of the variables ($a_{T,lm}$, $a_{E,lm}$, $a_{B,lm}$) together
with their cross-correlations.
For density (scalar) perturbations, the odd parity mode $B$ vanishes.
The power spectra can be written as (\cite{SeljakZ96})
\ba
C_{Tl} &\equiv& \langle |a_{T,lm}|^2 \rangle= (4\pi)^2 \int \frac{dk}{k}\,
A_S^2(k)\, \left| \Delta_{Tl} (k, \tau=\tau_0) \right|^2, \nonumber\\
C_{El} &\equiv &\langle |a_{E,lm}|^2 \rangle= (4\pi)^2 \int \frac{dk}{k}\,
A_S^2(k)\, \left| \Delta_{El} (k, \tau=\tau_0) \right|^2, \nonumber\\
C_{Cl} &\equiv &\langle a_{T,lm}^* a_{E,lm} \rangle= (4\pi)^2 \int \frac{dk}{k}\,
A_S^2(k)\, \Delta_{Tl} (k, \tau=\tau_0) \,
\Delta_{El} (k, \tau=\tau_0),
\ea
where $\Delta_{T,El} (k, \tau=\tau_0)$ are integrals over 
conformal time $\tau$ of the sources which generate the CMB temperature
fluctuations and polarization.

The assumption of random-phase density fluctuations does not necessarily imply
that the errors on the $C_{T,E,Cl}$ and the $P(k_q)$ are
Gaussian-distributed, although the Central Limit
Theorem allows us to make this approximation.  Following
Jungman \etal\ (1996), Goldberg \&
Strauss (1998), Tegmark (1997), and Zaldarriaga, Spergel, \& Seljak
(1997), 
we can then write the Fisher matrix for the combined MAP and SDSS data as: 
\be
\label{eq:fisher}
F_{ij}=\sum_l \sum_{X,Y} \frac{\partial C_{Xl}}{\partial s_i}\,
\mbox{Cov}^{-1}(C_{Xl}, C_{Yl}) \frac{\partial C_{Yl}}{\partial s_j}
+ \sum_q \frac{1}{\sigma^2_{P_q}}\,
\frac{\partial P_G(k_q)}{\partial s_i}\,\frac{\partial P_G(k_q)}{\partial s_j},
\ee
where $X,Y=T,E,C$. $ \mbox{Cov}^{-1}(C_{Xl}, C_{Yl})$ is the inverse
of the covariance matrix $\mbox{Cov}(C_{Xl}, C_{Yl})=\langle \Delta C_{Xl}
\Delta C_{Yl} \rangle$. For each $l$, one has to invert the covariance matrix 
and sum over $X$ and $Y$.
$\sigma_{P_q}$ are the standard errors in the measurement of $P(k_q)$
(equation~\ref{eq:sigmaP} below). 
We compute the $C_{Xl}$ and $P(k)$ using the 
CMBFAST Boltzmann code by Uro\v{s} Seljak
and Matias Zaldarriaga (\cite{SeljakZ96}).
We calculate those derivatives which cannot be found analytically
using two-sided finite differences.  We take step sizes of 
2\% of its model value for $h$, 
5\% of the model value of $\Omega_{m}$ for $\Omega_{\Lambda}$,
and 5\% of the model value of each parameter for the other parameters.

The covariance matrix of CMB power spectrum estimators has
diagonal elements given approximately by (\cite{Seljak96,ZS97,Kamion97}): 
\ba
\mbox{Cov}(C_{Tl}^2)&=&\frac{2}{(2l+1)f_{sky}}\, 
\left( C_{Tl}+ w_T^{-1} B_l^{-2}\right)^2, \nonumber \\
\mbox{Cov}(C_{El}^2)&=&\frac{2}{(2l+1)f_{sky}}\, 
\left( C_{El}+ w_P^{-1} B_l^{-2}\right)^2, \nonumber \\
\mbox{Cov}(C_{Cl}^2)&=&\frac{1}{(2l+1)f_{sky}}\, 
\left[ C_{Cl}^2+ \left( C_{Tl}+ w_T^{-1} B_l^{-2}\right)
\left( C_{El}+ w_P^{-1} B_l^{-2}\right) \right],
\ea
while the off-diagonal matrix elements are:
\ba
\mbox{Cov}(C_{Tl}C_{El})&=&\frac{2}{(2l+1)f_{sky}}\, C_{Cl}^2, \nonumber \\
\mbox{Cov}(C_{Tl}C_{Cl})&=&\frac{2}{(2l+1)f_{sky}}\, C_{Cl}
\left( C_{Tl}+ w_T^{-1} B_l^{-2}\right), \nonumber \\
\mbox{Cov}(C_{El}C_{Cl})&=&\frac{2}{(2l+1)f_{sky}}\, C_{Cl}
\left( C_{El}+ w_P^{-1} B_l^{-2}\right),
\ea
where $f_{sky}$ is the fraction of sky which is mapped.
We take $f_{sky}=0.8$ for MAP.
We have defined $w_{(T,P)}^{-1}\equiv\sigma_{(T,P)}^2\theta_{fwhm,c}^2$, 
where $\sigma_{T,P}$ is the noise per pixel in the temperature
and the polarization measurements, and 
$\theta_{fwhm}$ is the full width at half maximum of the beam in
radians. 
The quantity $B_l^2 \equiv \exp\left[-\left(0.425 \, 
\theta_{fwhm}\, l\right)^2\right]$ is the 
window function of  
the Gaussian beam. 

For MAP, both temperature and polarization data are obtained from the same
experiment by adding and differencing the two polarization
states, hence $\sigma_T^2=\sigma_P^2/2$, and the noise and
polarization signals are uncorrelated. 
MAP's resolution at its three highest frequency channels 
(40, 60, and 90$\,$GHz) is anticipated to be  
$\theta_{fwhm}=0.47^{\circ}$,
$0.35^{\circ}$, and $0.21^{\circ}$ respectively, with the corresponding 
$\sigma_T$=27$\,\mu$K, 35$\,\mu$K, and 35$\,\mu$K for its two
year nominal mission (Bennett \etal\ 1997; {\tt http://map.gsfc.nasa.gov}).
We do a weighted average in the three channels $c$, yielding
\ba
w_T  &=& \sum_c w_T^c= \sum_c \frac{1}{\sigma_{T,c}^2 
\theta_{fwhm,c}^2}=\frac{1}{(0.10\,\mu {\rm K})^2}, \nonumber \\
w_P &=& w_T/2,\nonumber \\
B_l^2 &=& \frac{\sum_c w_T^c B_{l,c}^2}{\sum_c w_T^c}
=\frac{\sum_c e^{-l^2 \sigma_{b,c}^2} /\left( \sigma_{T,c}^2\theta_{fwhm,c}^2
\right)}{\sum_c 1/ \left( \sigma_{T,c}^2\theta_{fwhm,c}^2 \right)}.
\ea
In our analysis, we ignore foreground contamination.  Current estimates
(Efstathiou \& Bond 1998; Refregier \etal\ 1998) suggest that foregrounds
will not significantly affect MAP temperature data.  Foreground
contamination, however, may significantly limit analyses
of polarization data, particularly at low $l$ (Keating \etal\ 1998).
Thus, we report
our results both for analyses that use only the temperature data
and for analyses that use both the temperature and polarization
data.

\subsection{Galaxy Data}

We take the SDSS standard error in the measured galaxy power spectrum
from Feldman, Kaiser, \& Peacock (1994).  For the power spectrum determined
over a volume in $k$-space of $V_k$: 
\be
\label{eq:dPk}
\frac{\sigma_{P}}{P_G(k)}=
\left( \frac{(2\pi)^3 \int d^3 {\bf r} \, \overline{n}^4({\bf r})
\psi^4 ({\bf r}) \left[ 1+\frac{1}{\overline{n}({\bf r})P_G(k)}
\right]^2 }
{V_k \left[ \int  d^3 {\bf r}\, \overline{n}^2({\bf r})\psi^2 ({\bf r}) 
\right]^2 }\right)^{1/2},
\label{eq:sigmaP}
\ee
where $\overline{n}({\bf r})$ is the selection function,
and $\psi ({\bf r})$ is the weighting function. 
For spherical shells,
$V_{k_q}=\frac{4}{3}\pi(k_q^3 -k^3_{q-1})$, where again, the spacing
between the $k_q$ is $2\pi/R$.  We choose $R=1466\,h^{-1}\,$Mpc for
the SDSS, following Goldberg \& Strauss (1998). 
We use the SDSS selection function calculated by Goldberg \& Strauss
(1998) using a mock SDSS redshift sample created by D. Weinberg 
(cf., Cole \etal\ 1998). 
We take the weighting function $\psi(\bfr) = 1$ throughout. 
Equation (\ref{eq:dPk}) assumes that the power estimates in each $k$
band are uncorrelated.  Goldberg \& Strauss (1998) show that this
assumption causes at most a 10\% underestimation of the true errors of
parameters.

\section{Determination of the Cosmological Parameters}

In this section, we discuss how well the cosmological parameters
can be determined by combining the MAP and SDSS data.  Although we
consider a general primordial power spectrum in Sec.~4, we here
restrict ourselves to the simpler case of a power-law model for the
primordial power spectrum generated by inflation, $k\,A_S^2(k) \propto
k^{n_S}$.  
We consider $h$, $\Omega_{\Lambda}$, $\Omega_b$, $\tau_{ri}$, 
$n_S$, $\Omega_m$, $C_2\equiv |a_{T,2m}|^2$ (which sets the CMB
normalization), and $b_{\it eff}$ as free parameters.  In Sec.~4, we
will see how our determination of these quantities worsens once we
give up a priori constraints on $A_S^2(k)$. 

The Fisher matrix depends on the true underlying power spectrum.  We
consider here three models, standard cold dark matter model (SCDM),
the CDM model with cosmological constant ($\Lambda$CDM), and open CDM (OCDM).
The model values of the cosmological parameters are shown in Table 1.
The $C_2$ values in Table 1 are obtained by normalizing
the models to the COBE four-year data, following \cite{Bunn&White}.
We take the galaxy bias parameter to be $b=1/\sigma_8$,
with $\sigma_8$ derived using the model $C_2$ value, and using the
standard value for the observed fluctuations of galaxies within
8$\,h^{-1}$Mpc spheres (\cite{DavisPeebles83}).  
Note that as we vary parameters to compute the
Fisher matrix, we keep $\Omega_m+\Omega_{\Lambda}=1$ for
the SCDM and $\Lambda$CDM models, because these models are flat by
definition. In the open CDM model, we allow for the simultaneous 
determination of $\Omega_m$ and $\Omega_{\Lambda}$.
\begin{table}[htb]
\caption{Model values of cosmological parameters}
\begin{center}
\begin{tabular}{ccccccccc}
\hline\hline
     & $h$  & $\Omega_{\Lambda}$ &$\Omega_b$& $\tau_{ri}$&$n_S$&$\Omega_m$& 
	$C_2$ &  $b=1/\sigma_8$ \\ 
\hline
SCDM &0.5 & 0 		       &  0.05       &  0.05       &1  &  1 & 
$1.139\times 10^{-10}$ &	0.83\\
$\Lambda$CDM &0.65 & 0.7       &  0.06       &  0.1        &1  &  0.3 & 
$1.455\times 10^{-10}$ & 1.29 \\
OCDM &0.65 & 0 		       &  0.06       &  0.05       &1  &  0.4& 
$1.607\times 10^{-10}$ & 1.72 \\
\hline
\end{tabular}
\end{center}
\end{table}

In the linear theory of mass density fluctuations, the bias factor $b$
is degenerate with the CMB normalization $C_2$.
Nonlinear effects in the present-day galaxy power spectrum break this degeneracy.
We include nonlinear effects following Peacock \& Dodds (1996).
Given a linear power spectrum $P_{L}(k_L)$,
the effective local power law index can be defined as
\be
n_{\it eff}(k_L)= n_L(k_L)= \left.\frac{d\ln P_L}{d\ln k} \right|_{k=k_L/2}.
\ee
The nonlinear power spectrum $P_{NL}(k)$
is related to the linear power spectrum by
\be
\Delta^2_{NL}(k_{NL})= f_{NL}\left(\Delta^2_{L}(k_L), \Omega \right),
\ee
where $\Delta^2_{L}(k_L)\equiv k_L^3 P_{L}(k_L)/(2\pi^2)$,
and $\Delta^2_{NL}\equiv k^3 P_{NL}(k)/(2\pi^2)$.
\be
f_{NL}(x, \Omega)=x \left[ \frac{1+B\gamma x+ [Ax]^{\alpha\gamma} }
{1+\left( [Ax]^{\alpha} g^3(\Omega)/[Vx^{1/2}] \right)^{\gamma} }
\right]^{1/\gamma}
\ee
with
\ba
A&=&0.482\, (1+n_{\it eff}/3)^{-0.947}, \nonumber \\
B&=&0.226\, (1+n_{\it eff}/3)^{-1.778}, \nonumber \\
\alpha&=&3.310\, (1+n_{\it eff}/3)^{-0.244}, \nonumber \\
\gamma&=&0.862\, (1+n_{\it eff}/3)^{-0.287}, \nonumber \\
V&=&11.55\, (1+n_{\it eff}/3)^{-0.423},
\ea
and (\cite{Carroll92})
\be
g(\Omega)= \frac{5}{2}\, \Omega_m \left[\Omega_m^{4/7}-\Omega_{\Lambda}
+(1+\Omega_m/2) (1+ \Omega_{\Lambda}/70 ) \right]^{-1}.
\ee
The nonlinear wavenumber is related to the linear wavenumber by
\be
k_{NL}=k_{L}\,[1+\Delta^2_{NL}(k_{NL})]^{1/3}.
\label{eq:kNL-def}
\ee
Note that although our consideration of nonlinear effects is adequate 
in the context of this paper, it is inaccurate in several ways.
First, even though we include nonlinear effects in $P(k)$,
we still use Kaiser's linear relation for redshift effects.
Proper consideration of redshift distortions on small scales 
(\cite{FisherNusser,Hatton97}) will lead to an effective bias $b_{\it eff}$ 
which depends on $k$ as well as the cosmological parameters $b$ and
$\Omega_m$.  Second, we have explicitly assumed that the true bias $b$
in real space is exactly linear and constant with scale; there must be
scale-dependence and non-linearities, especially at small scales,
where the density fluctuations are large (cf., Dekel \& Lahav 1998;
Blanton \etal\ 1998). 
Third, as the strength of the baryon-induced wiggles in the power
spectrum is proportional to $\Omega_b/\Omega_m$, they are more
pronounced in the $\Lambda$CDM and OCDM models. 
The Peacock \& Dodds (1996)
non-linear formulae only apply to {\it smooth\/} linear power spectra;
we therefore smooth the linear power spectra of the 
$\Lambda$CDM and OCDM models (using 
the analytical formulae in \cite{Eisen97})
before applying the non-linear formulae. Since we are discarding
the information contained in the baryon-induced wiggles, 
we obtain overestimates of the true errors.
Finally, non-linear effects will 
cause mode-mixing
between estimates of the power spectrum we have assumed here to be
independent (cf., Jain \& Bertschinger 1994).  These effects are
difficult to include in the Fisher matrix formalism, and we do not
attempt to do so here.  We carry out all calculations in this section
to two limits in the $k_q$ of the galaxy power spectrum,  
$k_{max} = 0.1\,h\,\rm Mpc^{-1}$ and $k_{max} = 0.5\,h\,\rm Mpc^{-1}$.
The Peacock \& Dodds (1996) formalism indicates that for the models
considered here, non-linear effects become significant for values of
$k$ somewhat larger than $0.1\,h \rm Mpc^{-1}$.  So the former value
of $k_{max}$
corresponds to a scale where non-linear effects are
negligible, while
the latter corresponds to a
scale $2\,\pi/k \approx 12 h^{-1}\rm Mpc$ where nonlinear effects are
becoming strong.  
It is work for the future to check the validity of
our calculations in this non-linear regime. 

To calculate the Fisher matrix, we need the derivatives of
the $C_{Xl}$'s ($X$ denotes $T$, $E$, $C$) and $P_G(k)=b_{\it eff}^2 P_{NL}(k)$ 
with respect to the cosmological parameters.
The derivatives with respect to $C_2$ and $b_{\it eff}$ can
be found analytically:
\ba
\label{eq:dCl,PG/dC2,db}
\frac{\partial C_{Xl}}{\partial \ln C_2} &=& C_{Xl}, \nonumber\\
\frac{\partial C_{Xl}}{\partial \ln b_{\it eff}} &=& 0, \nonumber\\
\frac{\partial P_G(k)}{\partial \ln C_2} &=& P_G(k)\, 
\frac{\partial\,\ln\,f_{NL}(x,\Omega)}{\partial\,\ln\,x}, \nonumber\\
\frac{\partial P_G(k)}{\partial \ln b_{\it eff}} &=& 2\, P_G(k),
\ea
where $x\equiv\Delta_L^2(k_L)$, and $k_L$ is related to $k$ via
equation~(\ref{eq:kNL-def}). In the OCDM model,
we take both $\Omega_{\Lambda}$ and $\Omega_m$ to be free parameters. Hence
\be
\label{eq:dPGdOL}
\frac{\partial P_G(k)}{\partial \Omega_{\Lambda}} = P_G(k)\,
\left[ \left.\frac{\partial\,\ln\,f_{NL}(x,\Omega)}{\partial\Omega_{\Lambda}}
\right|_{x} + \left.
\frac{\partial\,\ln\,f_{NL}(x,\Omega)}{\partial\,\ln\,x}
\right|_{\Omega_{\Lambda}} \cdot 
\frac{\partial\,\ln\,P_L(k_L)}{\partial\Omega_{\Lambda}}\right],
\ee
where $P_L(k_L)$ depends on $\Omega_{\Lambda}$ only through an overall
normalization for the $k$ range relevant to redshift surveys
(on superhorizon scales, $P_L(k_L)$ has a $k$-dependent 
$\Omega_{\Lambda}$ dependence in the OCDM model).
The derivatives with respect to the other cosmological parameters
are calculated numerically by finite differences.

Tables 2, 3, and 4 show the Fisher Matrix 1-$\sigma$ error bars 
for the cosmological parameters, $\Delta s_i=\sqrt{({\bf F}^{-1})_{ii} }$
for SCDM, $\Lambda$CDM, and OCDM respectively.  
For each parameter, we
list the errors for six cases: the MAP temperature data taken alone,
the MAP temperature data with SDSS data to $k_{max} = 0.1\,h\rm
Mpc^{-1}$, the MAP temperature data with SDSS data to $k_{max} = 0.5\,h\rm
Mpc^{-1}$, and the corresponding results when MAP polarization data
are included as well. 
\begin{table}[htb]
\caption{Fisher Matrix 1-$\sigma$ error bars for the cosmological 
parameters of SCDM.}
\begin{center}
\begin{tabular}{ccccccc}
\hline\hline
 1-$\sigma$ error    &  
MAP$^T$& 
$\begin{array}{ll}\mbox{MAP}^T+\\ \mbox{SDSS}^1 \end{array}$ &
$\begin{array}{ll}\mbox{MAP}^T+\\ \mbox{SDSS}^2 \end{array}$ &
MAP$^{(T+P)}$ & 
$\begin{array}{ll}\mbox{MAP}^{(T+P)}+\\ \mbox{SDSS}^1 \end{array}$ &
$\begin{array}{ll}\mbox{MAP}^{(T+P)}+\\ \mbox{SDSS}^2 \end{array}$ 
\\ 
\hline
$\Delta \ln C_2$    	 &  0.17      	& 0.097 &  0.016           & 
0.065	 & 0.053	& 0.016\\

$\Delta \ln h$ 	     	 &  0.052 & 0.041  & 0.021		&
0.033	& 0.028		&0.018	\\

$\Delta \Omega_{\Lambda}$& 0.15 & 0.11 & 0.057	 	&	
0.091	& 0.076	& 0.047       \\

$\Delta \ln (\Omega_b h^2)$  & 0.028	&0.024 	    & 0.019		&
0.020	& 0.019	& 0.017	\\

$\Delta \ln \tau_{ri}$	 &2.4		&1.5	 &0.46		&
0.39		 	 &0.38		& 0.30	\\

$\Delta \ln n_S$  		 &0.017	& 0.014& 0.0081	&      
0.0085	& 0.0076	&0.0059		\\

$\Delta \ln b_{\it eff}$		 &  		& 0.064 & 0.028	&
			 &0.063	&0.022		\\
\hline
\end{tabular}
\tablecomments{``T'' denotes temperature only, ``(T+P)'' denotes
temperature plus polarization for MAP data.
``1'' and ``2'' denote the SDSS cutoff $k_{max}=0.1\,h\rm Mpc^{-1}$, 
and $0.5\,h\rm Mpc^{-1}$ respectively.}
\end{center}
\end{table}

\begin{table}[htb]
\caption{Fisher Matrix 1-$\sigma$ error bars for the cosmological 
parameters of $\Lambda$CDM.}
\begin{center}
\begin{tabular}{ccccccc}
\hline\hline
 1-$\sigma$ error    &  
MAP$^T$& 
$\begin{array}{ll}\mbox{MAP}^T+\\ \mbox{SDSS}^1 \end{array}$ &
$\begin{array}{ll}\mbox{MAP}^T+\\ \mbox{SDSS}^2 \end{array}$ &
MAP$^{(T+P)}$ & 
$\begin{array}{ll}\mbox{MAP}^{(T+P)}+\\ \mbox{SDSS}^1 \end{array}$ &
$\begin{array}{ll}\mbox{MAP}^{(T+P)}+\\ \mbox{SDSS}^2 \end{array}$ 
\\ 
\hline
$\Delta \ln C_2$    	 &  0.21      	& 0.15 &  0.010            & 
			0.043	 & 0.043	& 0.010\\

$\Delta \ln h$ 	     	 &  0.066& 0.048 & 0.025 &
		0.032	&0.030		&0.021	\\

$\Delta \ln \Omega_{\Lambda}$& 0.076 & 0.055 & 0.027	 	&	
0.037	&0.035	& 0.023       \\

$\Delta \ln (\Omega_b h^2)$  & 0.044	& 0.034 & 0.022		&
0.021	&0.021	& 0.018	\\

$\Delta \ln \tau_{ri}$	 &1.33		&0.95	 &0.18	&
0.18		 	 &0.18		& 0.13	\\

$\Delta \ln n_S$  		 &0.035	& 0.026 & 0.014		&      
0.014	& 0.013	& 0.011		\\

$\Delta \ln b_{\it eff}$		 &  	& 0.076 & 0.022		&
			 &0.038	& 0.021		\\
\hline
\end{tabular}
\end{center}
\end{table}

\begin{table}[htb]
\caption{Fisher Matrix 1-$\sigma$ error bars for the cosmological 
parameters of OCDM.}
\begin{center}
\begin{tabular}{ccccccc}
\hline\hline
 1-$\sigma$ error    &  
MAP$^T$& 
$\begin{array}{ll}\mbox{MAP}^T+\\ \mbox{SDSS}^1 \end{array}$ &
$\begin{array}{ll}\mbox{MAP}^T+\\ \mbox{SDSS}^2 \end{array}$ &
MAP$^{(T+P)}$ & 
$\begin{array}{ll}\mbox{MAP}^{(T+P)}+\\ \mbox{SDSS}^1 \end{array}$ &
$\begin{array}{ll}\mbox{MAP}^{(T+P)}+\\ \mbox{SDSS}^2 \end{array}$ 
\\ 
\hline
$\Delta \ln C_2$    	 &  0.13      	& 0.10 &  0.034            & 
0.081	 & 0.064	& 0.033\\

$\Delta \ln h$ 	     	 &  0.11 & 0.085 & 0.018		&
0.093	& 0.069		&0.015	\\

$\Delta \ln \Omega_m$& 0.13 & 0.099& 0.011	 	&	
0.13	&0.083	& 0.010       \\

$\Delta \Omega_{\Lambda}$& 0.088 & 0.060 & 0.010	 	&	
0.085	& 0.055	& 0.0097       \\

$\Delta \ln (\Omega_b h^2)$  & 0.041	& 0.033 	  & 0.017	&
0.038	&0.031	& 0.017	\\

$\Delta \ln \tau_{ri}$	 &1.5		&1.5	 &0.46	&
0.57		 	 &0.57		& 0.37	\\

$\Delta \ln n_S$  	& 0.026	& 0.021 & 0.0079		&      
0.023	&0.020	& 0.0076		\\

$\Delta \ln b_{\it eff}$	&  	& 0.14 & 0.029		&
			 & 0.14	& 0.026		\\
\hline
\end{tabular}
\end{center}
\end{table}

The peak features (called ``Doppler peaks'', ``acoustic peaks'', or 
``Sakharov peaks'') in the CMB angular power spectrum $C_{Tl}$ are due to 
acoustic oscillations in the baryon-photon fluid at the time of last 
scattering of the CMB photons.
As mentioned above, the amplitude of these peaks is determined by
$\Omega_b h^2$ and $\Omega_m h^2$, while their location is determined 
by the geometry of the universe (roughly $\Omega_m + \Omega_\Lambda$
in a nearly flat universe), hence there is a near
degeneracy in one dimension of the parameter space of
$h$, $\Omega_b$, $\Omega_{\Lambda}$ and $\Omega_m$ (Bond \etal\ 1994;
Zaldarriaga, Spergel \& Seljak 1997; \cite{HD97}). For models
with only six or seven free parameters, this degeneracy is not severe
(see Table 2); however, it is significant if we consider models
with many more free parameters (e.g., a variable spectral index, or massive neutrinos).

Large scale structure observations are sensitive to an orthogonal set of
parameters.
The turnover in the matter power spectrum is set by 
the horizon size at matter-radiation equality
(e.g., Peebles 1993).  Thus,
SDSS will make an accurate measurement of $\Omega_m h$. 
The Silk damping scale, and the acoustic oscillations also affect 
the transfer function, so that $\Omega_b h$ can also be determined 
from the SDSS observations (Goldberg \& Strauss 1998).
Thus combining the MAP and SDSS data allows us to do two important
things.  First, to the extent that parameters derived from the two
independent data sets are consistent, we obtain a powerful check of the
correctness of the physical model which ties the two together:
gravitational instability theory, the assumption that the fluctuations
are adiabatic, and the linear biasing paradigm.  
This is illustrated in Figure 1, which shows the 68.3\%
and 95.4\% confidence contours  
in the $\Omega_b$-$h$ (a) and $\Omega_m$-$h$ (b) planes. 
The dotted lines are for MAP temperature data only, 
the dashed lines are for SDSS data only ($k_{max}=0.5\,h$Mpc$^{-1}$), and
the solid lines are for the combined MAP temperature and SDSS data.   
The SCDM model is assumed. 
Second, the joint analysis of the two independent data sets
breaks some of the degeneracies inherent 
in the MAP data taken alone and improves the accuracy
in the determination of parameters.
Figure 2 shows the 68.3\% and 95.4\% confidence contours  
in the $\Omega_{\Lambda}$-$\Omega_m$ plane in the OCDM model,
with the same line types as in Figure 1.
Note that using MAP data alone, there is degeneracy between $\Omega_{m}$ 
and $\Omega_{\Lambda}$; using SDSS data alone, $\Omega_{\Lambda}$
can not be determined because it is degenerate with $C_2$
(see equation~\ref{eq:dPGdOL}).
The combined data give a much better determination of $\Omega_{m}$
and $\Omega_{\Lambda}$ than does either the MAP data 
or the SDSS data alone.

The CMB temperature spectrum is also approximately degenerate in its
dependence on the reionization optical depth $\tau_{ri}$ and the CMB
normalization $C_2$, because reionization suppresses the spectrum by a
factor of $\exp(-2\tau_{ri})$ on small scales relative to the large
scales. Figure 3 shows the confidence contours
in the $\tau_{ri}$-$C_2$ plane, with the same line types as in Figure 1.  
Combining MAP data with SDSS data
dramatically breaks the degeneracy between $\tau_{ri}$ and $C_2$ by
adding information on the small scales.

Figures 1-3 show MAP temperature data only, because the extent
to which MAP polarization data can be used may be limited by 
foregrounds (see Section 5). Also, if we allow a cutoff of
$k_{max}=0.5\,h\,$Mpc$^{-1}$ (as we do for the rest of this paper), 
adding MAP polarization data
to the combined MAP and SDSS data does not have a significant impact
on the accuracy of the determination of cosmological parameters
(see Tables 2-4).

Note that the errors of the parameters from combined MAP and SDSS data
are very sensitive to the SDSS cutoff $k_{max}$. This is as expected,
since the small scale information from the SDSS is determined
by its cutoff $k_{max}$.  Indeed, in most cases, the galaxy data to
$k_{max} = 0.1\, h\rm Mpc^{-1}$ gives only a moderate improvement to the
constraints the MAP data alone give; much of the improvement comes
from the non-linear galaxy regime.

To summarize, when MAP temperature data is considered alone, there is
strong degeneracy between the overall amplitude of the matter power spectrum
$C_2$ and the reionization optical depth $\tau_{ri}$,
and between the matter density fraction $\Omega_m$ and the density fraction
contribution from cosmological constant $\Omega_{\Lambda}$.
Combining MAP and SDSS data dramatically breaks the degeneracy
between $C_2$ and $\tau_{ri}$, and reduces the 
degeneracy between $\Omega_m$ and $\Omega_{\Lambda}$,
leading to a significant improvement on the determination 
of $C_2$, $\tau_{ri}$, $\Omega_m$, and $\Omega_{\Lambda}$, enabling us to measure
the effective bias between the matter power spectrum and
the galaxy redshift power spectrum $b_{eff}$.
The accuracy in the determination of other cosmological parameters
is also improved by combining MAP and SDSS data, but to a less
impressive extent.
In the next section, we study the measurement 
of the primordial power spectrum itself, which can only be achieved by
using the combined MAP and SDSS data.

\section{Measurement of the Primordial Power Spectrum}

In the previous section, we considered a power-law primordial power
spectrum, and derived the constraints on cosmological models that the
combination of MAP and SDSS data will give.  We now consider a more
general primordial power spectrum; as we discussed in the Introduction,
various inflationary models make a range of predictions for the form
of the primordial power spectrum. 
We parameterize the primordial power spectrum $A_S^2(k)$ as
\be
A_S^2(k) = \left\{ \begin{array}{ll}
a_1, \hskip 1cm \mbox{for}\,\, k<k_1, \\
a_i, \hskip 1cm \mbox{for}\,\, k_{i-1}<k<k_i, \,\,i>1\end{array} \right.
\ee
where $k_1=0.001\,h$Mpc$^{-1}$, and $k_i$ ($i=2,3,...,20$) are equally
spaced in log$\,k$ from $k_1$ to $k=0.5h\,$Mpc$^{-1}$.
The $a_i$ ($i=1,2,...,20$) are taken to be independent variables.
To compute errors in the determination of the $a_i$, we need to know
their true values, as well as true values of the cosmological
parameters.
We first consider the case that the cosmological parameters are known
a priori; we consider the realistic case in which we know neither the
primordial power spectrum nor the cosmological parameters later in
this section.  For definiteness, we adopt a model in which the true primordial power
spectrum is $A_S^2(k)=1$, and the cosmological parameters are those of
SCDM (see Table 1). 


In the Fisher matrix approach we have adopted in this paper, we
need to compute the derivatives of the observables with respect
to the parameters, which now include the values of the steps $a_i$
which make up the primordial power spectrum.
The derivatives of the $C_l$'s with respect to $a_i$ are taken to be
finite differences with step-size of 5\% the model value of $a_i$.
The derivatives of the galaxy power spectrum with respect to 
$a_i$ are given by
\be
\frac{\partial \ln P_G(k)}{\partial a_i}=
\left\{ \begin{array}{ll}
\frac{\partial\, \ln\,f_{NL}(x)}{\partial \,\ln \, x} \, \frac{1}{a_i},
\hskip 1cm \mbox{for}\,\, k<k_1\,\, (i=1), \mbox{or}\,\, 
 k_{i-1}<k<k_i \,\,(i>1),\\
0,  \hskip 1cm \mbox{otherwise,} \end{array} \right.
\ee
where $x$ and $f_{NL}(x)$ are the same as in Equation (\ref{eq:dCl,PG/dC2,db}).

In this section, we take the cutoff wavenumber of the SDSS
to be $k=0.5$Mpc$^{-1}$ throughout. Also, when we consider
the combined MAP and SDSS data, only the temperature data from
MAP is used; adding the polarization data does not have a significant effect 
on errors from the combined MAP and SDSS data, while it is of great 
interest to consider the SDSS data as an {\it alternative} to the
MAP polarization data in constraining inflationary models.

Figure 4 shows the accuracy in the determination of the bin amplitudes
of the primordial power spectrum, $k\,A_S^2(k)$; the 1-$\sigma$ error bars
are shown for MAP temperature data only (a), 
SDSS data only (b), and MAP temperature and SDSS data combined (c).
Assuming the cosmological parameters are known,
for 20 bins in $\log \,k$ ($k\leq 0.5h$Mpc$^{-1}$),
the primordial power spectrum can be determined to around 16\% 
accuracy for $k\sim 0.01\,h\,$Mpc$^{-1}$, and to
around 1\% accuracy for $k\sim 0.1\, h\,$Mpc$^{-1}$ (see Figure 4(c))
by combining MAP temperature and SDSS data.

Not surprisingly, the determination of the primordial power spectrum
bin amplitudes does depend on the a priori knowledge of the
cosmological parameters.  We now consider the realistic case in which
the cosmological parameters are not known a priori; in particular, in
addition to the $a_i$, we also solve for the four 
parameters $h$, $\Omega_{\Lambda}$, $\Omega_b$, and $\tau_{ri}$.
The primordial power spectrum bin amplitudes
and the cosmological parameters can not be simultaneously determined
using MAP temperature data only, because $\tau_{ri}$ is degenerate
with the $a_i$ for $k\ga 0.01\,h\,$Mpc$^{-1}$.
However, polarization data removes this degeneracy.
Figure 5(a) shows the resulting error bars on the $a_i$ if both the 
temperature and polarization data from MAP are used; they are 
significantly larger than in Figure 4(a) where only temperature data
is used.

Figure 5(b) shows the effect of combining the MAP temperature data with
the SDSS data in the
analysis, adding the galaxy bias $b_{\it eff}$ as a fifth parameter. 
The uncertainty in the determination of the primordial power spectrum
bin amplitudes $a_i$ increases by a factor up to 3 on small scales 
relative to the
case in which the cosmological parameters are known {\it a priori}
(Figure 4(c)). This difference is indistinguishable in the figures,
because it enters through error bars that are only of the order of 
a few percent.
The errors in the $a_i$ are not significantly increased by the inclusion
of cosmological parameters in the parameter estimation,
because most of the information on the cosmological parameters
comes from small angular scales, where the power spectrum can be determined
to great accuracy from the SDSS data.
Comparison of Figure 5(a) and Figure 5(b) illustrates the critical importance
of combining the MAP data with the SDSS data in 
the determination of the primordial power spectrum bin amplitudes.

Note that our choice of 20 bins in log$\,k$ to parameterize $A_S^2(k)$
is somewhat arbitrary. Obviously,
the errors on the bin amplitudes would 
increase if we were to increase the number of bins.
When the cosmological parameters are estimated together with
the $A_S^2(k)$ bin amplitudes $a_i$, we expect the 
covariance between the cosmological parameters and the $a_i$'s to
increase rapidly with smaller bin size, as the primordial power
spectrum becomes flexible enough to mimic the Doppler peaks and other
features of the $C_l$. 

To illustrate how the errors on the cosmological parameters change
as they are estimated jointly with primordial power spectrum bin
amplitudes using the combined MAP temperature and SDSS data, we show in
Table 5 the Fisher matrix 1-$\sigma$ error bars for
two groups of cosmological parameters,
($h$, $\Omega_b$, $\tau_{ri}$, $b_{\it eff}$)
and ($h$, $\Omega_{\Lambda}$, $\Omega_b$, $\tau_{ri}$, $b_{\it eff}$),
when they are estimated 
jointly within each group and when each group is simultaneously
estimated with 20 independent bin amplitudes of the primordial power
spectrum. 
\begin{table}[htb]
\caption{Fisher Matrix 1-$\sigma$ error bars for the cosmological 
parameters.}
\begin{center}
\begin{tabular}{ccccc}
\hline\hline
 1-$\sigma$ error    &4 parameters& 
$\begin{array}{ll}
\mbox{4 parameters \& 20 bin}\\ \mbox{amplitudes of } A_S^2(k) \end{array}$ 
&5 parameters&
$\begin{array}{ll}
\mbox{5 parameters \& 20 bin}\\ \mbox{amplitudes of } A_S^2(k) \end{array}$ 
\\ 
\hline
$\Delta \ln h$ 	     &  0.0048& 0.0062
  		     & 0.017 & 0.025	\\

$\Delta \Omega_{\Lambda}$&     & 
			& 0.046	& 0.064 \\

$\Delta \ln (\Omega_b h^2)$  & 0.016 & 0.039 	
			    & 0.016 & 0.046 \\

$\Delta \ln \tau_{ri}$	 & 0.056	& 0.19	
			 & 0.12	& 0.27	\\

$\Delta \ln b_{\it eff}$	 &  0.0052 & 0.0094
			 & 0.012	&  0.017       \\
\hline
\end{tabular}
\end{center}
\end{table}
The cosmological parameters can still be determined to
impressive accuracy even when they are estimated simultaneously 
with 20 independent bin amplitudes of the primordial power spectrum!
Of course, we do not include the overall amplitude 
of the power spectrum $C_2$ as a free parameter, as it is degenerate
with the $a_i$ by definition. 

Figures 4 and 5 assume that the true primordial power spectrum
is $A_S^2(k)=1$. For the purpose of illustration, we can 
choose our model primordial power spectrum $A_S^2(k)$ to 
have broken scale-invariance based on the
Holman \etal~(1991) and Adams \etal~(1997) models.
We write
\be
\label{eq:As2}
A_S^2(k)= \left\{ \begin{array}{lll}
&&1, \hskip 1cm k<0.01\, h \mbox{Mpc}^{-1}, \\
&& A+B\,k, \hskip 1cm
0.01\, h \mbox{Mpc}^{-1} <k < 0.1\, h \mbox{Mpc}^{-1}, \\
&&0.25\,k^{-0.2}, \hskip 1cm k>0.1\, h \mbox{Mpc}^{-1},
\end{array}\right.
\ee
where $A$ and $B$ are chosen such that $A_S^2(k)$ is continuous.
Figure 6 shows the primordial power spectrum of Equation (\ref{eq:As2})
with 1-$\sigma$ error bars for the combined MAP temperature and SDSS data.
Clearly, the accuracy in the determination of $A_S^2(k)$ is 
not sensitive to the assumed underlying model;
the errors on $A_S^2(k)$ bin amplitudes do not change significantly
with model where the errors are larger than of order a few percent.
  
There have been recent claims of possible observational detection of
features in the matter power spectrum on the scale
of $k\sim 0.1\, h$Mpc$^{-1}$ (\cite{Peacock97}). 
Combining MAP and SDSS data
will enable us to confirm such features, and to determine whether they
are intrinsic to the primordial power spectrum, independent of
specific inflationary models. 
Indeed, it would be straightforward to translate our bin amplitude errors of
the primordial power spectrum into constraints on existing inflationary models;
our errors are small enough to be very interesting.
However, it is perhaps more exciting to anticipate that the
measurement of the primordial power spectrum
in wavenumber bins from the combined MAP and SDSS data will provide a lead
to the correct inflationary model.

\section{Summary}

Cosmic microwave background experiments measure fluctuations in the
curvature of space at the surface of last scattering.  If
the primordial fluctuations are adiabatic, then MAP  will be a sensitive
probe of the geometry of the universe, the matter/photon
ratio ($\Omega_m h^2$), the baryon/photon ratio ($\Omega_b
h^2$) and the primordial power spectrum for $k < 0.2\,h\rm Mpc^{-1}$.

Observations of large scale structure measure the large scale
distribution of galaxies.  If the primordial fluctuations
are adiabatic, then SDSS will be a sensitive probe of
the primordial power spectrum for  $k > 0.02\,h\rm Mpc^{-1}$ and
$\Omega_m h$, which determines the horizon 
size at matter-radiation equality and thus the shape of the processed
power spectrum. SDSS will also be a powerful probe
of the nature of the dark matter (Hu \etal\ 1997).
Because of the large number of galaxies with detailed photometric
and spectroscopic data in the SDSS, we can
determine the power spectrum and bias factor for several
independent samples of galaxies, defined by morphology, color, or
spectral characteristics.  To the extent to which these independent samples
yield consistent power spectra, we can determine the range
in $k$ over which the linear bias model is valid.

The combination of the SDSS and the MAP data will be
a powerful test of our basic cosmological models.  Both
experiments will accurately determine
the amplitude of density fluctuations
at $0.2 > k > 0.02\, h\rm Mpc^{-1}$. Will
there be a set of cosmological parameters and primordial power spectrum
that is consistent with both sets of observations?  If so,
then we will have tested the gravitational structure
formation paradigm, our interpretation of the primordial
fluctuations as adiabatic, and the linear biasing paradigm.
If not, this conflict will likely lead us to a deeper
understanding of the origin of structure in the universe.

In particular, from the combination of the MAP and SDSS data, we can obtain
a measurement of the primordial power spectrum without reference to
any specific inflationary model, assuming that the primordial
fluctuations are adiabatic. 
If the values of the cosmological parameters that best fit the data
are close to what we expect, but the primordial power spectrum differs
significantly from the predictions of the current inflationary models,
then it will be an indication of new physics in the early universe,
and it will provide a solid starting point for building new 
inflationary models. 

If there is a theory consistent with both data sets, then
the combination of the two observations will break the
degeneracies inherent in each individual observation.  Cosmic microwave
background observations by themselves measure a combination of
the amplitude of the primordial power spectrum and the optical
depth.  Large scale structure observations measure the product
of the current matter power spectrum and the bias parameter.  In
linear theory, the offset between the power spectrum
as determined by the two sets of observations  determines
a combination of the bias parameter and the optical depth
of the universe.  Because of non-linear effects on the spectrum,
this offset is scale-dependent, this will enable us to independently
measure the bias parameter and the optical depth of the universe.
The combination of the two experiments will also improve
our ability to determine other cosmological parameters,
in particular, the matter density and the cosmological constant.  

Measurements of the cosmic microwave background polarization can 
provide an independent measure of the optical
depth of the universe $\tau_{ri}$ (Zaldarriaga, Spergel \& Seljak 1997).
However, emission from both polarized galactic dust and
synchrotron emission (Keating \etal\ 1998)
may swamp
the primordial polarization fluctuations at large
angular scales.  These foregrounds may significantly limit our ability
to extract useful cosmological information from
polarization measurements.  
Hence it is important that we can use the SDSS data as an alternative
and independent aid to MAP in the determination of $\tau_{ri}$.

The Sloan Digital Sky Survey, and other large scale structure surveys,
will make other independent measurements that will probe
cosmological parameters and the primordial spectrum: redshift
distortions, cluster properties, small scale velocity
fields, and the evolution of structure.  These will provide
additional tests of the basic model and will further
enhance our ability to measure cosmological parameters.  

With this in mind, there are a number of improvements that could be
done in this analysis.  The redshift distortions could be measured
directly, and the effects could be included directly into the Fisher
matrix analysis (cf., de Laix \& Starkman 1998 and 
Hatton \& Cole 1998 for a discussion
of how well redshift-space distortions can be measured 
from the SDSS data).
Similarly, we could parameterize the effects of
galaxy and clustering evolution, and include these as parameters in
the analysis.  More challenging will be a proper accounting of
non-linear effects on small scales, in the power spectrum, the
redshift space distortions, and the bias (here assumed linear and
independent of scale).  The analyses in Tables~2-4 
show that going to $k_{max} =  0.5\,h\rm Mpc^{-1}$ gives us
particularly strong constraints on cosmological parameters, but we
have taken the non-linear effects into account in only a relatively
crude way in this paper.  It remains to be seen the extent to which
the uncertainties due to these inaccuracies strongly affect the error
bars we derive on the primordial power spectrum. 

The next few years will be a very exciting time in cosmology.

\section{Acknowledgements}

Y.W. acknowledges partial support from NSF grant AST94-19400.
DNS acknowledges support from the MAP/MIDEX project. MAS acknowledges
support from the Alfred P. Sloan Foundation, a fellowship from
Research Corporation, and NSF grant AST96-16901. 
We acknowledge the use of CMBFAST Boltzmann code by Uro\v{s} Seljak
and Matias Zaldarriaga. It is a pleasure for us to thank Matias
Zaldarriaga for many helpful clarifications concerning CMBFAST,
and Daniel Eisenstein and Wayne Hu for emphasizing the importance
of two-sided derivatives.

\clearpage

\setcounter{figure}{0}
\figcaption[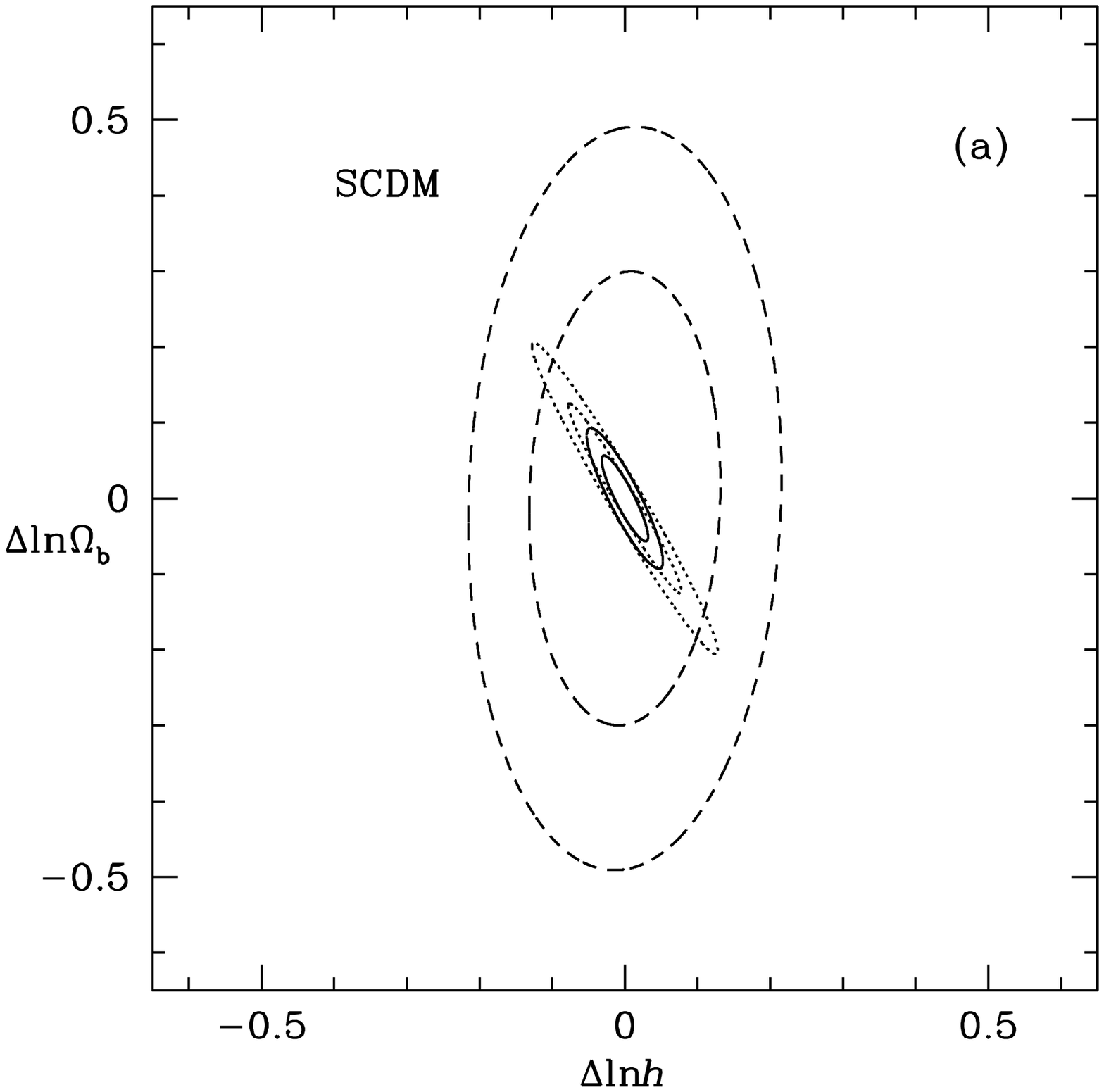]{The confidence contours (68.3\% and 95.4\%)
in the $\Omega_b$-$h$ (a) and $\Omega_m$-$h$ (b) planes assuming the SCDM model.
The dotted lines are for MAP temperature data only, 
the dashed lines are for SDSS data to $k_{max} = 0.5\,h \rm Mpc^{-1}$ 
only, and the solid lines are for the combined MAP temperature and SDSS data.}

\figcaption[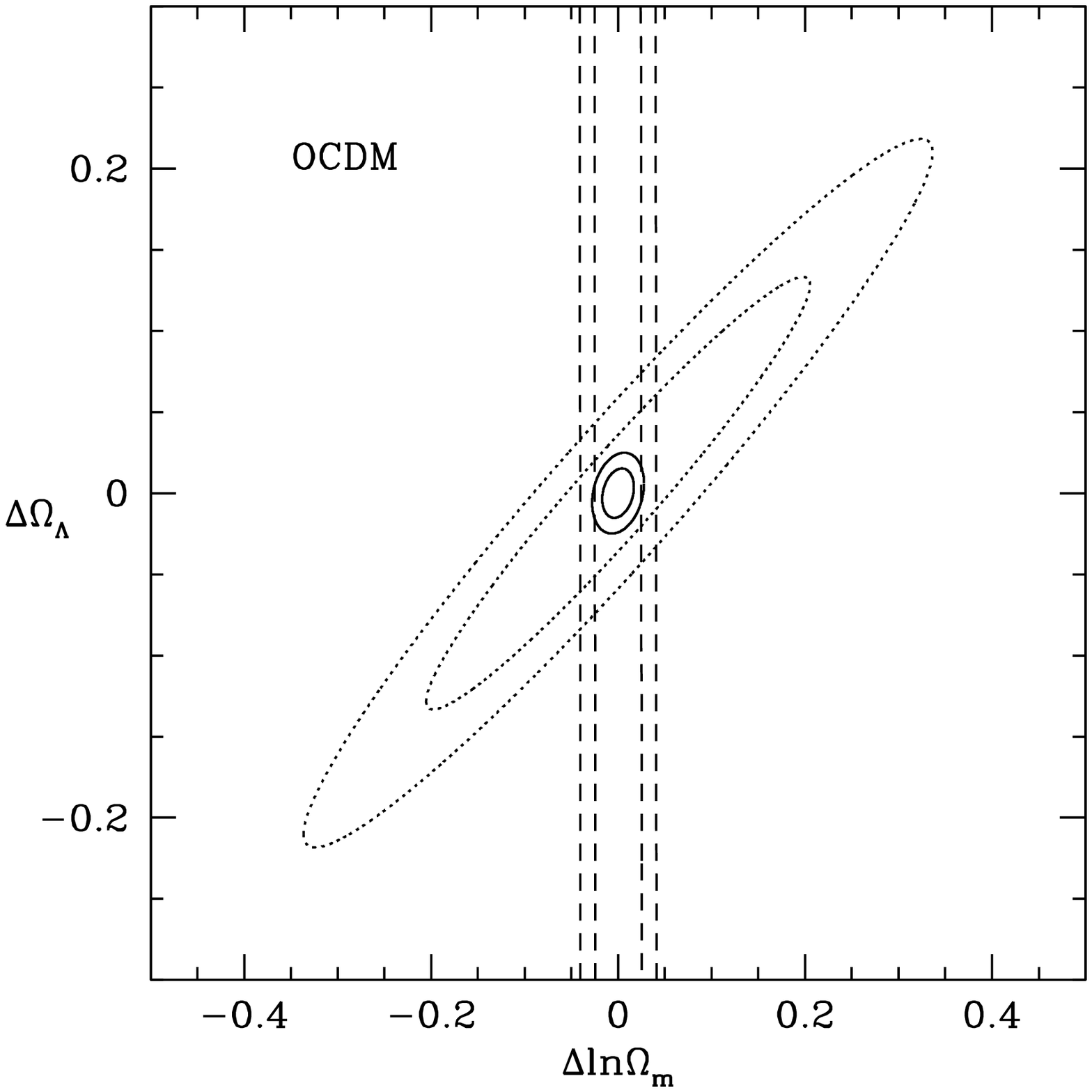]{The confidence contours (68.3\% and 95.4\%)
in the $\Omega_{\Lambda}$-$\Omega_m$ plane for the OCDM model.
The line types are the same as in Figure 1.
}

\figcaption[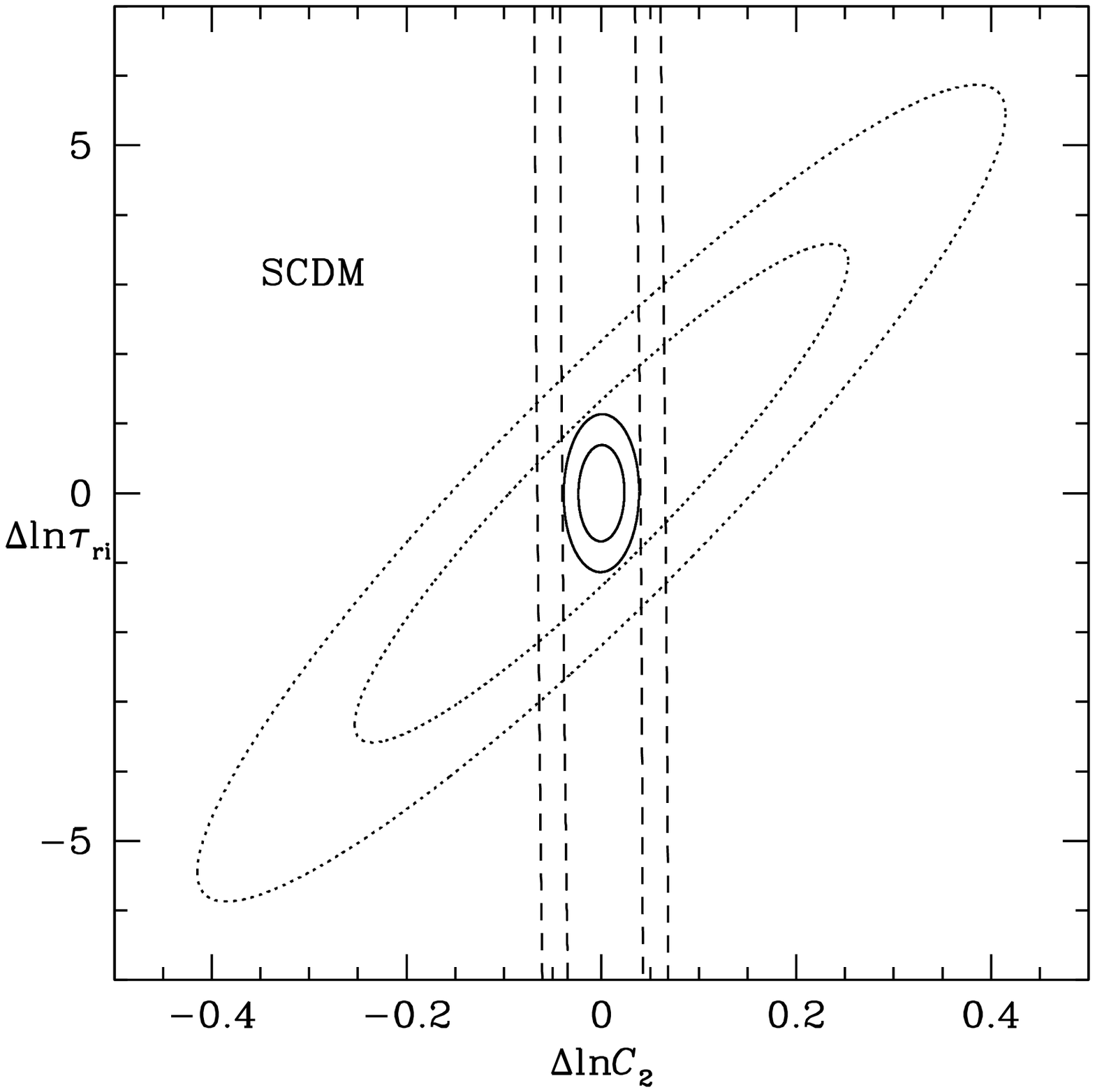]{The confidence contours (68.3\% and 95.4\%)
in the $\tau_{ri}$-$C_2$ plane for the SCDM model. 
The line types are the same as in Figure 1.
}

\figcaption[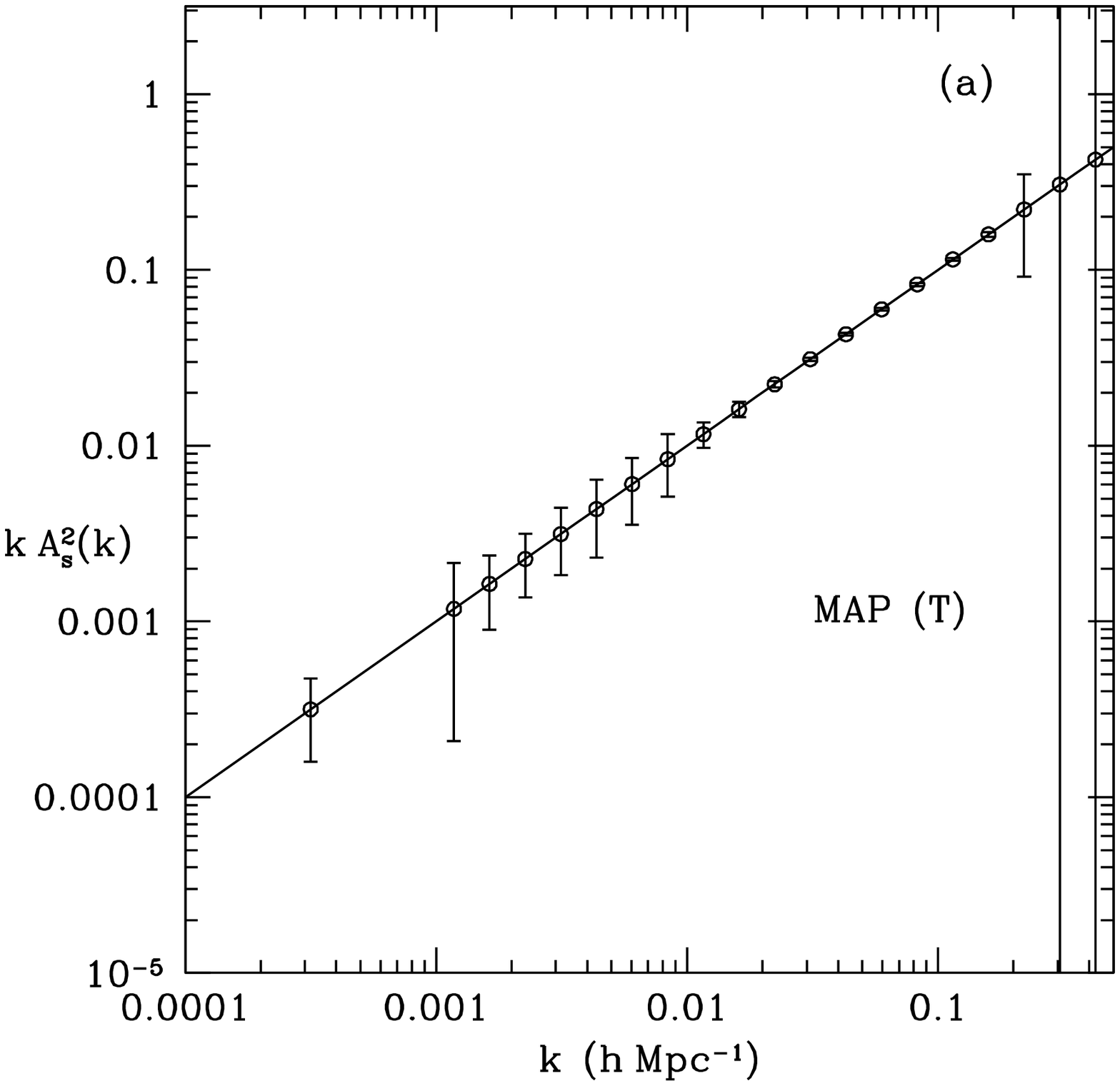]{The primordial power spectrum with 1-$\sigma$ 
error bars for MAP temperature data only (a), SDSS data only (b),
and MAP temperature and SDSS data combined (c),
assuming that the cosmological parameters are known to be 
that of the SCDM model (with $A_S^2(k)=1$) listed in Table 1.
The error bars in (b) which extend
to the edges of the figure are infinite, as these refer to
measurements of the power spectrum on scales larger than those which
SDSS probes. 
}

\figcaption[As2MAP4p.eps]{The primordial power spectrum with 1-$\sigma$ 
error bars.
(a) Combined MAP temperature and polarization data,  
with four cosmological parameters 
($h$, $\Omega_{\Lambda}$, $\Omega_b$, $\tau_{ri}$)
included in the parameter estimation;
(b) Combined MAP temperature and SDSS data, with
five cosmological parameters
($h$, $\Omega_{\Lambda}$, $\Omega_b$, $\tau_{ri}$, $b$) 
included in the parameter estimation.}

\figcaption[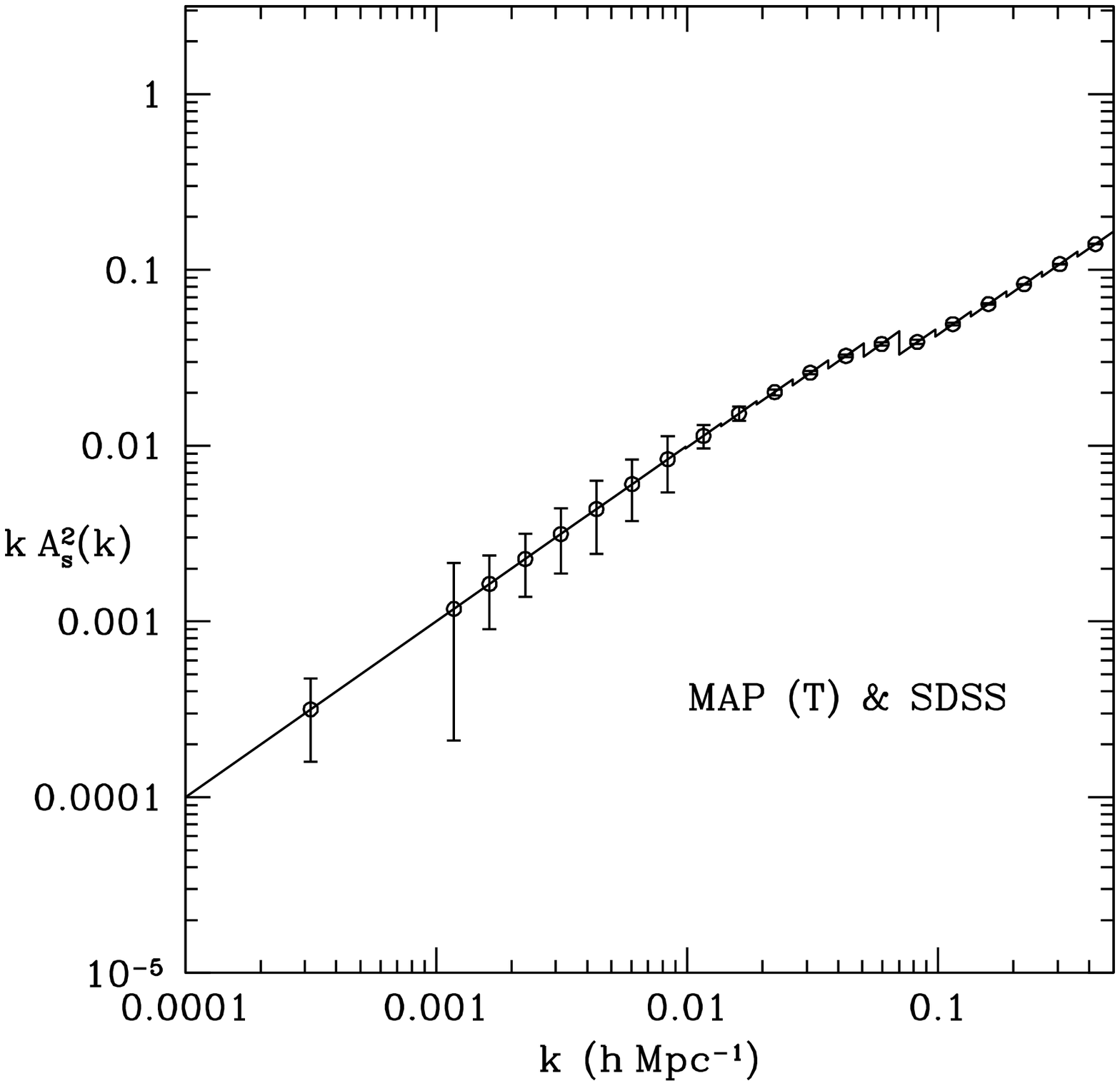]{The primordial power spectrum of Equation (\ref{eq:As2}) with 1-$\sigma$ error bars for the combined MAP temperature 
and SDSS data.
}
  
\clearpage

\setcounter{figure}{0}
\plotone{Obh.eps}
\figcaption[Obh.eps]{
(a) The confidence contours (68.3\% and 95.4\%)
in the $\Omega_b$-$h$ plane assuming the SCDM model.
The dotted lines are for MAP temperature data only, 
the dashed lines are for SDSS data to $k_{max} = 0.5\,h \rm Mpc^{-1}$ 
only, and the solid lines are for the combined MAP temperature and SDSS data.}
\setcounter{figure}{0}
\plotone{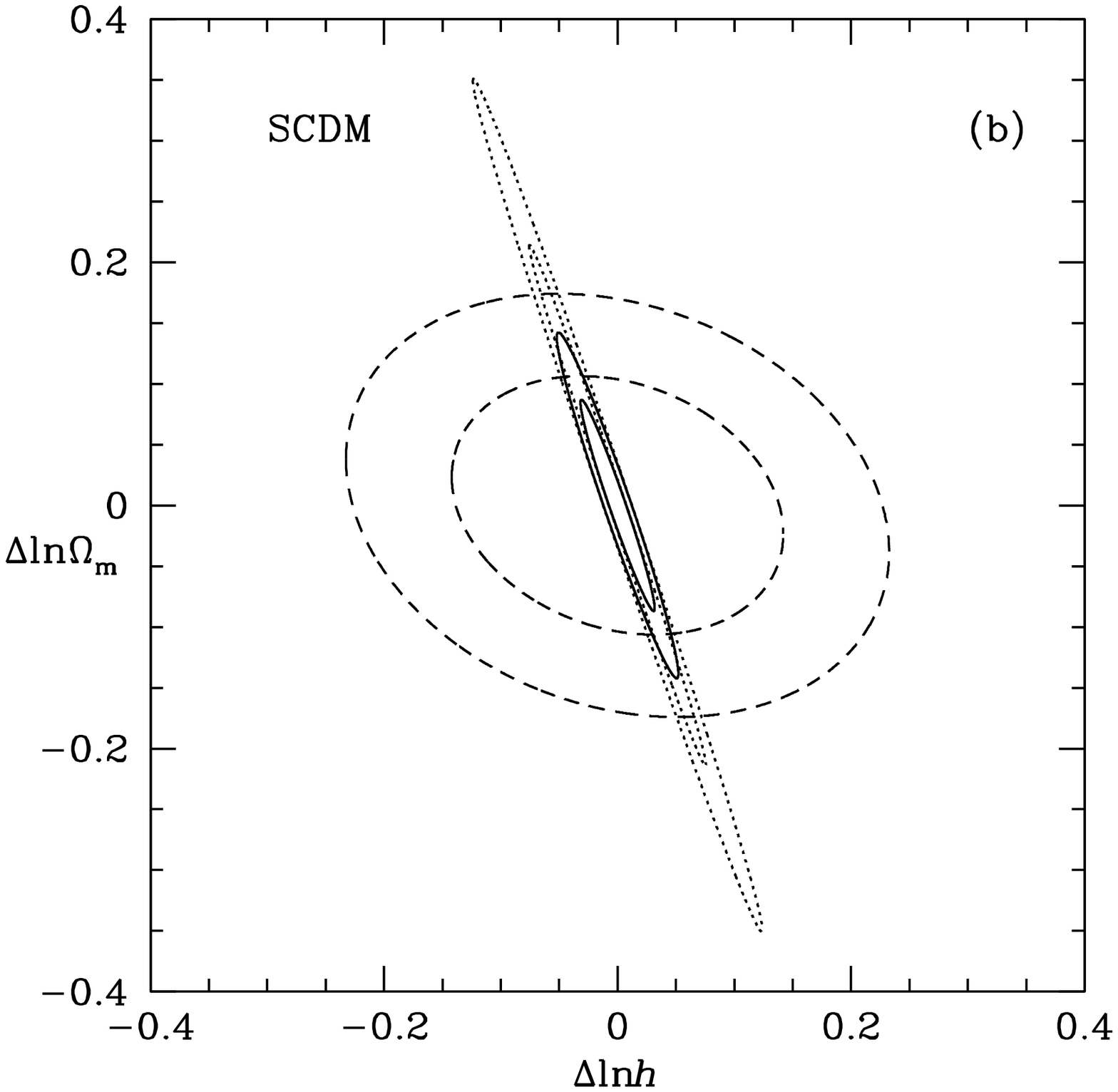}
\figcaption[Obh.eps]{
(b) The confidence contours (68.3\% and 95.4\%)
in the $\Omega_m$-$h$ plane.
The line types are the same as in (a). }

\plotone{OCDMOmOL.eps}
\figcaption[OCDMOmOL.eps]{
The confidence contours (68.3\% and 95.4\%)
in the $\Omega_{\Lambda}$-$\Omega_m$ plane for the OCDM model.
The line types are the same as in Figure 1.}

\plotone{tauC2.eps}
\figcaption[tauC2.eps]
{The confidence contours (68.3\% and 95.4\%)
in the $\tau_{ri}$-$C_2$ plane for the SCDM model. 
The line types are the same as in Figure 1.}

\plotone{As2MAP.eps}
\figcaption[As2MAP.eps]{
The primordial power spectrum with 1-$\sigma$ error bars, assuming that the cosmological parameters are known to be 
that of the SCDM model (with $A_S^2(k)=1$) listed in Table 1.
(a) MAP temperature data only.}

\setcounter{figure}{3}
\plotone{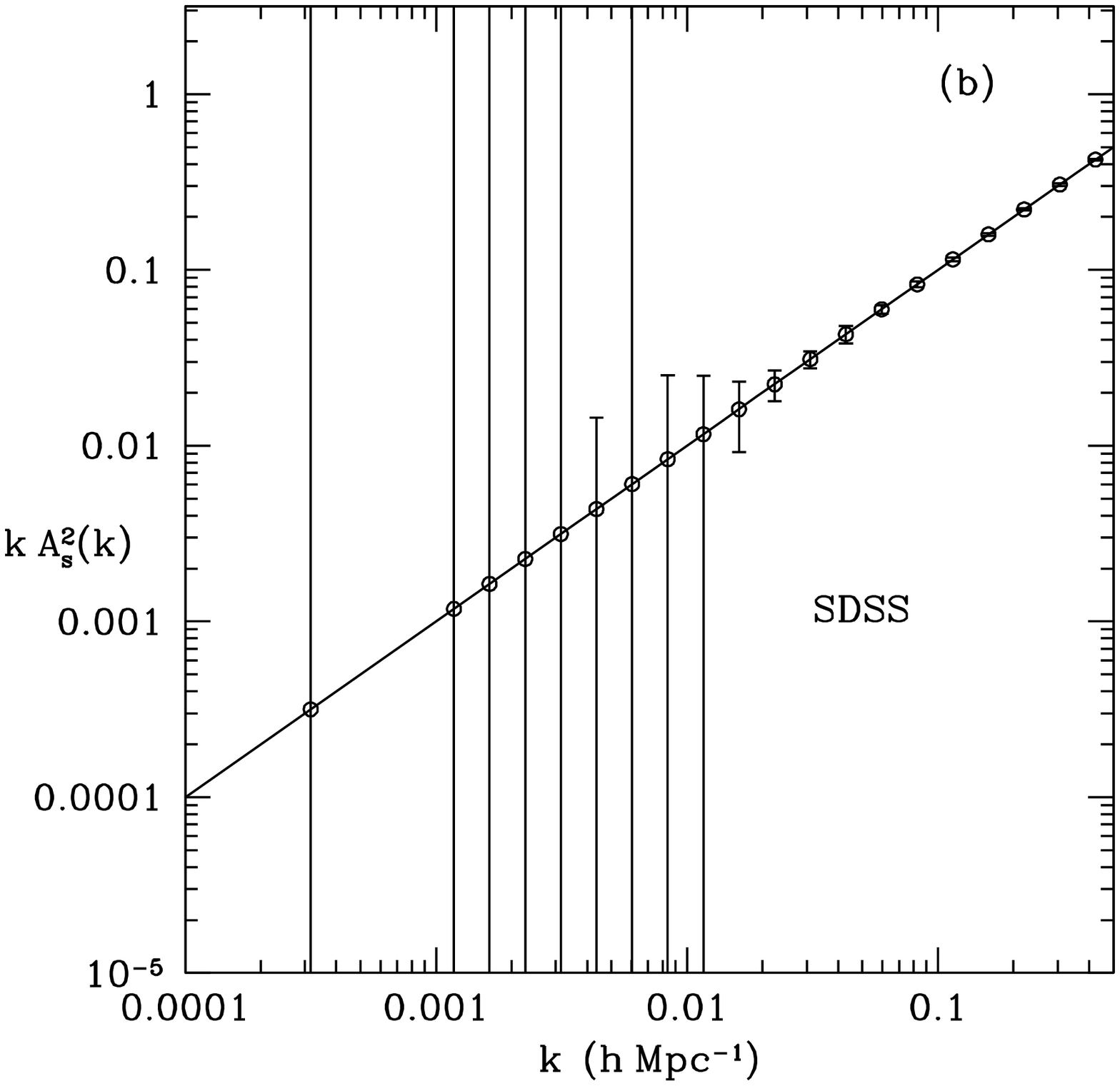}
\figcaption[As2SDSS.eps]{(b) SDSS data only.  The error bars which extend
to the edges of the figure are infinite, as these refer to
measurements of the power spectrum on scales larger than those which
SDSS probes. 
}

\setcounter{figure}{3}
\plotone{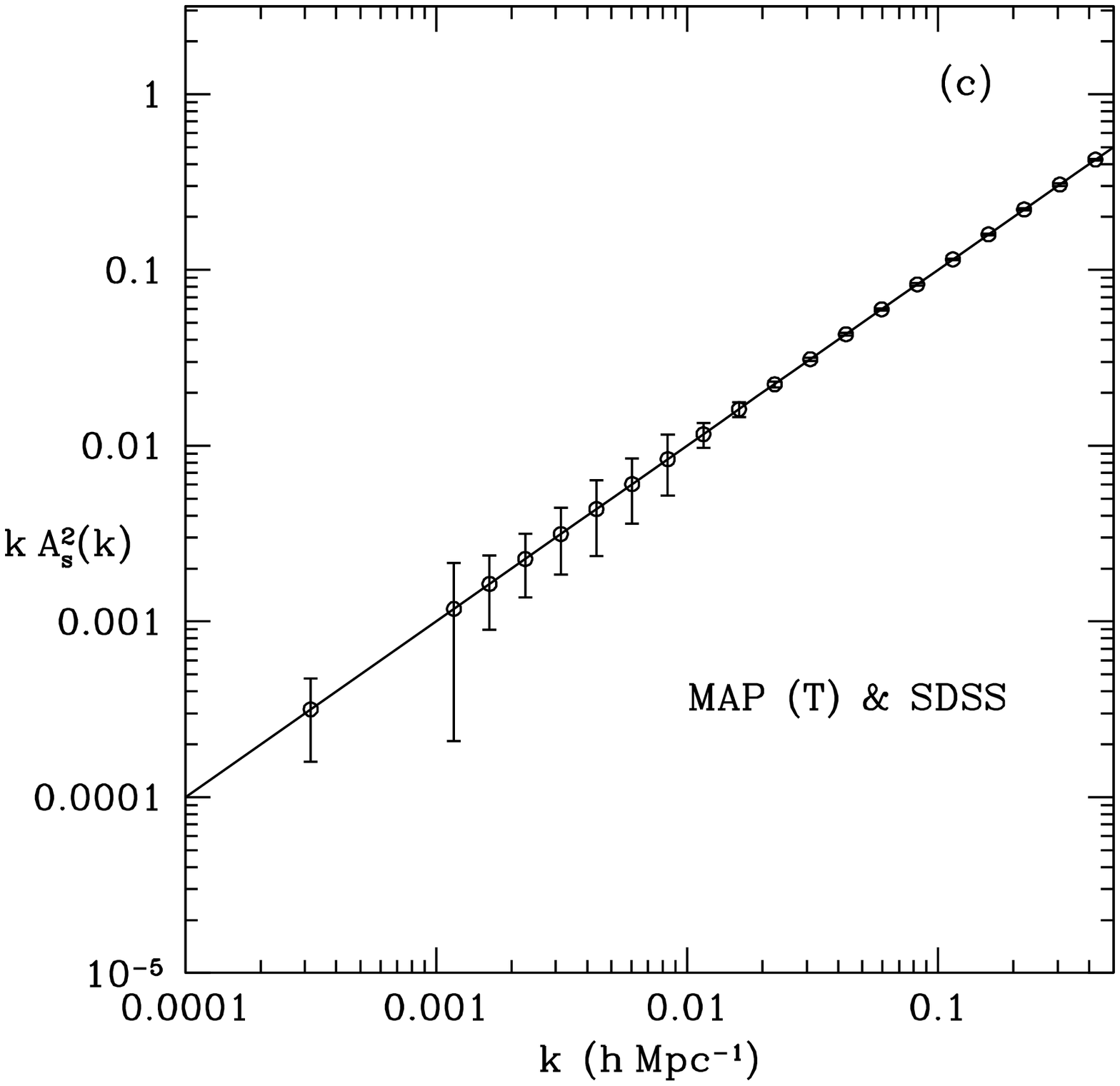}
\figcaption[As2MAPSDSS.eps]{(c) MAP temperature and SDSS data combined.} 

\setcounter{figure}{4}
\plotone{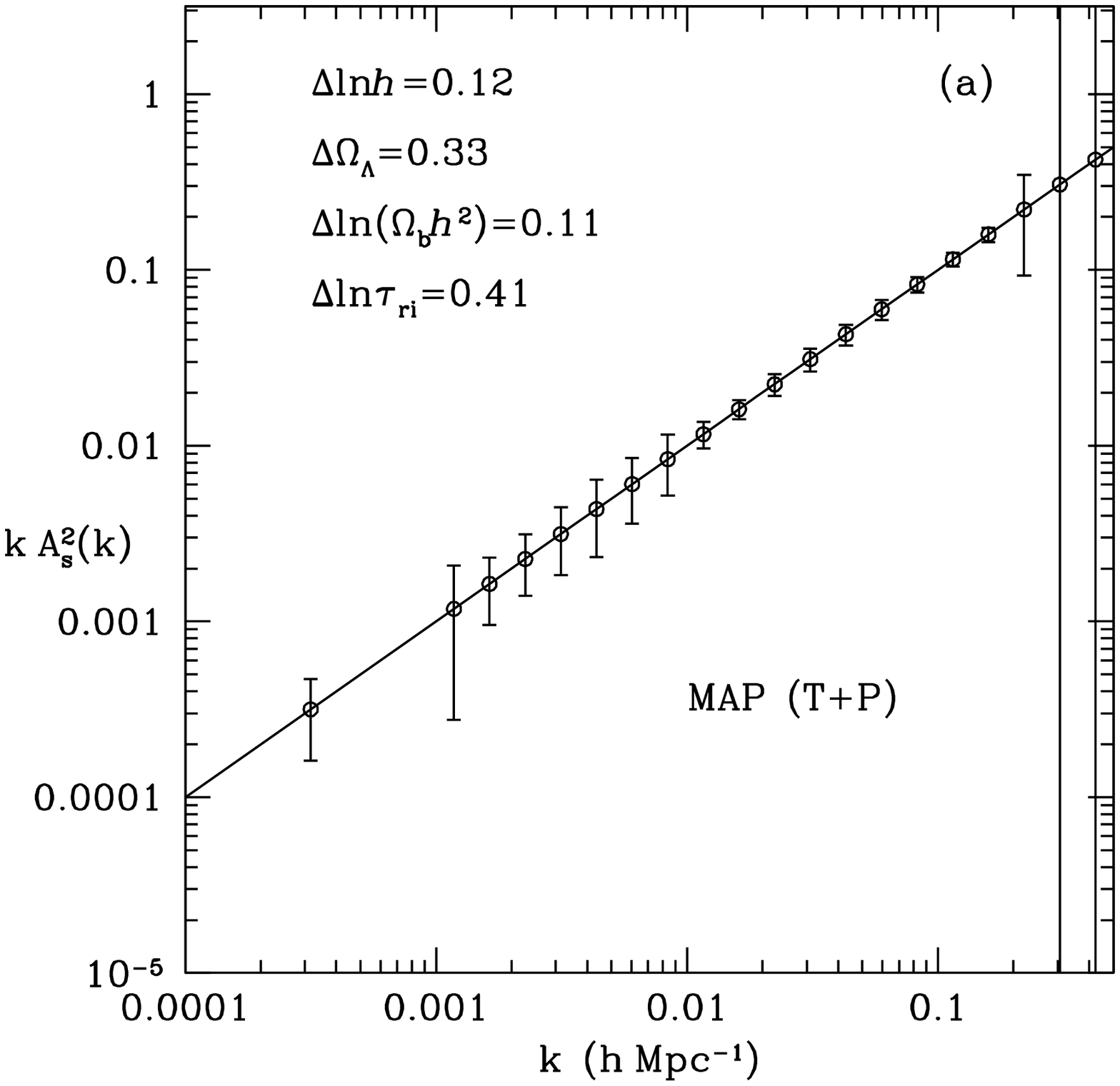}
\figcaption[As2MAP4p.eps]{
The primordial power spectrum with 1-$\sigma$ error bars.
(a) Combined MAP temperature and polarization data,  
with four cosmological parameters 
($h$, $\Omega_{\Lambda}$, $\Omega_b$, $\tau_{ri}$)
included in the parameter estimation.}

\setcounter{figure}{4}
\plotone{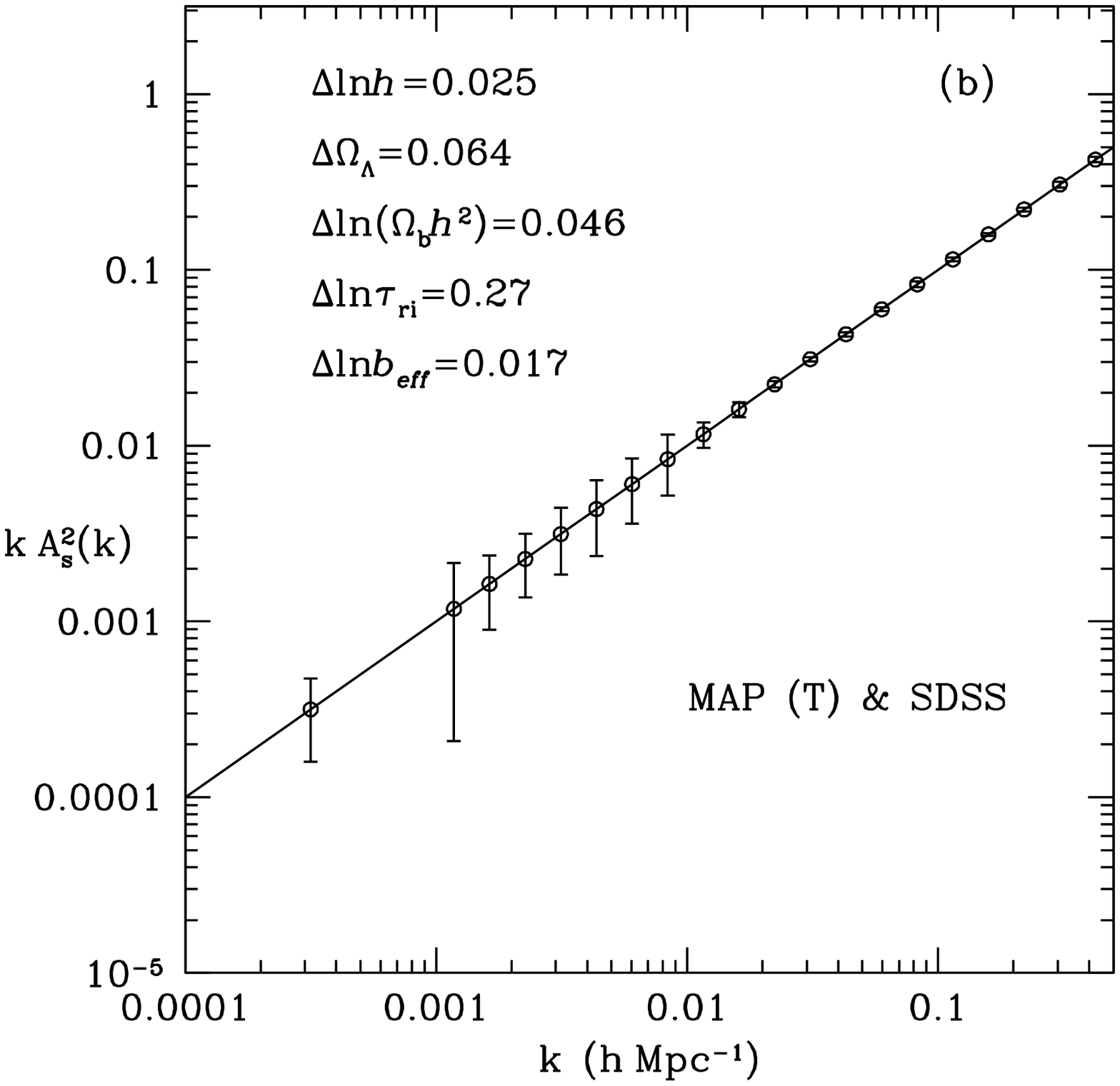}
\figcaption[As2MAPSDSS5p.eps]{
(b) Combined MAP temperature and SDSS data, with
five cosmological parameters
($h$, $\Omega_{\Lambda}$, $\Omega_b$, $\tau_{ri}$, $b$) 
included in the parameter estimation.}

\plotone{As2MAPSDSSb.eps}
\figcaption[As2MAPSDSSb.eps]{
The primordial power spectrum of Equation (\ref{eq:As2})
with 1-$\sigma$ error bars for the combined MAP temperature and SDSS data.}

\end{document}